\documentclass[twocolumn,showpacs,amsmath,amssymb,aps,prb,superscriptaddress]{revtex4-1}

\usepackage[]{verbatim}
\usepackage[]{color}
\usepackage[abs]{overpic}
\usepackage{rotating}
\usepackage{tabularx}

\usepackage{amsfonts}
\usepackage{amsmath}
\usepackage{amssymb}
\usepackage{bm}
\usepackage{graphicx}
\usepackage[breaklinks,colorlinks=true,citecolor=blue]{hyperref}
\usepackage[]{cleveref}
\usepackage{mathrsfs}
\usepackage[lofdepth,lotdepth,caption=false]{subfig}
\usepackage{varwidth}
\usepackage{wrapfig}
\usepackage{times}
\usepackage{longtable}
\usepackage{multirow}
\usepackage{tikz}
\usepackage{colortbl}

\usepackage{times}

\usepackage[]{color}
\usepackage{overpic}
\usepackage{rotating}
\usepackage{mathtools}
\usepackage{tabularx}

\definecolor{darkblue}{RGB}{0 60 120}
\definecolor{eggplant}{RGB}{190 10 150}
\definecolor{darkgray}{RGB}{70 70 70}
\definecolor{lightgray}{RGB}{80 80 80}
\definecolor{lightgray2}{RGB}{245 215 110}
\definecolor{lightgray3}{RGB}{255 0 0}

\graphicspath{{./}}

\begin{document}

\title{Spin-orbit-entangled nature of magnetic moments and Kitaev magnetism in layered halides $\alpha$-RuX$_3$ (X = Cl, Br, I)}

\author{Heung-Sik Kim}
\email{heungsikim@kangwon.ac.kr}
\affiliation{Department of Physics and Institute for Accelerator Science, Kangwon National University, Chuncheon 24341, Korea}

\begin{abstract}
$\alpha$-RuCl$_3$ has been extensively studied recently because of potential bond-dependent Kitaev magnetic exchange interactions and the resulting quantum spin liquid phase that can be realized therein. It has been known that the covalency between Ru 4$d$- and Cl $p$-orbitals is crucial to induce large Kitaev interactions in this compound, therefore replacing Cl into heavier halogen elements such as Br or I can be a promising way to promote the Kitaev interaction even further. In a timely manner, there have been reports on synthesis of $\alpha$-RuBr$_3$ and $\alpha$-RuI$_3$, which are expected to host the same spin-orbit-entangled orbitals and Kitaev exchange interactions with $\alpha$-RuCl$_3$. Here in this work we investigate electronic structures of $\alpha$-RuCl$_3$, $\alpha$-RuBr$_3$, and $\alpha$-RuI$_3$ in a comparative fashion, focusing on the cooperation of the spin-orbit coupling and on-site Coulomb repulsions to realize the spin-orbit-entangled pseudospin-1/2 at Ru sites. We further estimate magnetic exchange interactions of all three compounds, showing that $\alpha$-RuBr$_3$ can be promising candidates to realize Kitaev spin liquid phases in solid-state systems. 
\end{abstract}
\maketitle


\section{Introduction}
Since Kitaev's finding of an exactly solvable magnetic model on a two-dimensional honeycomb lattice\cite{kitaev2006anyons}, which promises a fault-tolerant quantum computation via realization of Majorana fermions with non-Abelian statistics, there have been a plethora of theoretical and experimental studies aiming to summon this Kitaev physics upon condensed matter systems. A general guiding principle has been suggested by the seminal work of Jackeli and Khaliullin\cite{jackeli2009mott}, followed by a number of theoretical and experimental studies searching for and evaluating the viability of the so-called Jackeli-Khaliullin mechanism in realistic systems\cite{Winter2017,KitaevReview}. A most promising candidate so far is $\alpha$-RuCl$_3$\cite{plumb2014alpha,hskim_RuCl3,jasears,luke2,luke,HSK2016,Banerjee,Do2017,Kasahara2018,Sears2020,Yokoi2021}, where half-quantized thermal Hall conductivity, considered to be a smoking-gun evidence for the presence of the Kitaev spin liquid (KSL) state, has been reported\cite{kitaev2006anyons,Kasahara2018,Yokoi2021}. 

An obstacle in reaching the KSL state in $\alpha$-RuCl$_3$ is the presence of magnetic exchange interactions, aside from Kitaev's, which induce static magnetism and break the KSL phase. Specifically the presence of the first- and third-nearest-neighbor Heisenberg interaction is detrimental for the Kitaev magnetism\cite{Winter2016}, which are found to be quite significant and hard to be completely removed due to the multi-orbital nature of candidate systems and the resulting various hopping processes\cite{rau2014generic}. Therefore, selective enhancement of relevant electron hopping channels contributing to the Kitaev's exchange interaction is essential for the realization of the KSL phase in realistic situations.

Recently there have been a couple of theoretical and experimental reports on $\alpha$-RuBr$_3$\cite{Salavati2019,Ersan2019,Imai2021} and $\alpha$-RuI$_3$\cite{Ersan2019,Ni2021,Zhang2021}, focusing on the possibility of promoting and realizing the KSL phase. As discussed in more detail in the following, the heavier ligand involves in stronger hybridization between Ru 4$d$- and ligand (Cl, Br, I) $p$-orbitals, which may lead to stronger Kitaev interactions. On the other hand, it may also induce unwanted enhancement of further-neighbor and even inter-layer Heisenberg interactions, which are not favorable to realize the KSL phases. Enhanced bandwidth due to larger $d$-$p$ hybridization can also disturb the formation of the local spin-orbital-entangled $j_{\rm eff}$ = 1/2 moments, a critical element of the Kitaev's exchange interactions. Therefore evaluation of such quantities via first-principles electronic structure calculations is crucial at this early stage of studies on $\alpha$-RuBr$_3$ and $\alpha$-RuI$_3$.

Hence in this work we focus on the electronic structure of $\alpha$-RuBr$_3$ and $\alpha$-RuI$_3$, in comparison to that of $\alpha$-RuCl$_3$, and the formation of the spin-orbital-entangled $j_{\rm eff} = 1/2$ moments in the compounds. We further estimate magnitudes of magnetic exchange interactions via employing first-principles Wannierization of Ru $t_{\rm 2g}$ bands, arguing that $\alpha$-RuBr$_3$ can be a promising platform to further investigate the Kitaev magnetism in condensed matter systems.

\section{Computational details}
\label{sec:detail}
For the optimizations of cell parameters and internal coordinates of all three compounds, we employed 
the Vienna {\it ab-initio} Simulation Package ({\sc vasp}), which uses the 
projector-augmented wave (PAW) basis set\cite{VASP1,VASP2}. 
500 eV of plane wave energy cutoff was used, and for
$k$-point sampling 7$\times$7$\times$3 $\Gamma$-centered grid 
were adopted. Effects of electron correlations in structural optimizations were considered via the Strongly Constrained and Appropriately Normed ({\sc scan}) semi-local functional\cite{SCAN}, which is parameter-free, on top of a revised Perdew-Burke-Ernzerhof (PBEsol)\cite{PBEsol} exchange-correlation functional. After structural optimizations, effects of atomic spin-orbit coupling (SOC) and on-cite Coulomb repulsions were studied via a linear-combinaion-of-pseudo-atomic-orbital basis code {\sc openmx}\cite{openmx,han2006n}, where projection onto the spin-orbit-entangled $j_{\rm eff} = 1/2$ and $3/2$ states and the Wannierization of Ru $t_{\rm 2g}$ bands were performed\cite{MLWF1,MLWF2,Weng2009}. Double zeta plus polarization (DZP) bases and 300 Ry of energy cutoff for real space integrations were employed for the {\sc openmx} calculations, and the effects of on-site Coulomb repulsions were incorporated via a simplified flavor of the rotationally-invariant DFT+$U$ formalism\cite{Dudarev} on top of the Perdew-Zunger parameterization for the local density approximation\cite{CA,PZ}. The value of $U_{\rm eff} \equiv U-J$ was chosen to be 2 eV, which has been considered to be a reasonable value\cite{hskim_RuCl3,HSK2016}. Computing electron hopping integrals using Wannier orbital method were done in the nonmagnetic phase without including SOC and on-site Coulomb interactions to avoid double-counting of such effects. 

\begin{figure}
  \centering
  \includegraphics[width=0.45 \textwidth]{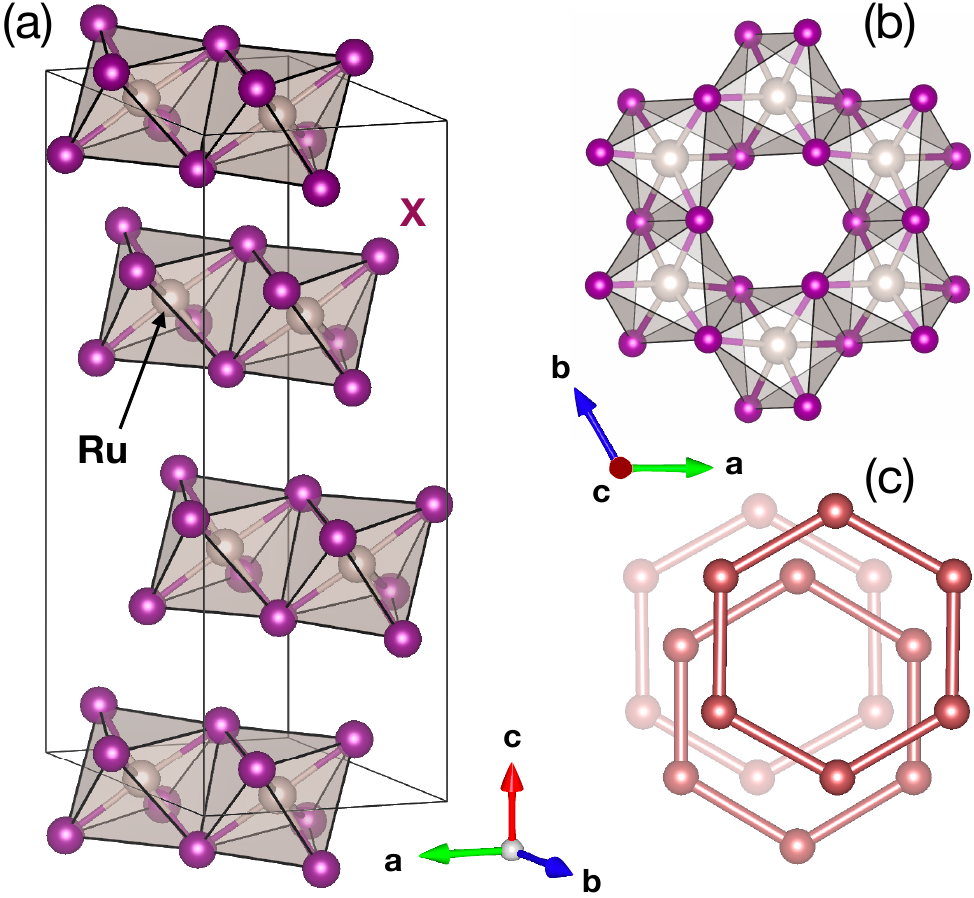}
  \caption{(Color online)
  (a) Crystal structure of $\alpha$-RuX$_3$ (X = Cl, Br, I) with $P3_112$ space group. 
  Solid lines depict the trigonal unit cell. 
  (b) A top-down view of the single RuX$_3$ layer, where the stacking of neighboring 
  Ru honeycomb layers are shown in (c).
  }
  \label{fig:struct}
\end{figure}

\section{Crystal structure}

Fig.~\ref{fig:struct} shows the model crystal structure we chose for $\alpha$-RuX$_3$ (X = Cl, Br, I) in this study (Space group: $P3_112$). There have been reports of different space group symmetries and stacking patterns for these compounds\cite{Imai2021,Zhang2021}, but our choice of the $P3_112$ stacking does not become a serious issue because interlayer interactions are fairly small in all three compounds\cite{HSK2016}. Also note that, while the zigzag-type antiferromagnetic order (which requires larger in-plane periodicity) has been reported as the ground state for $\alpha$-RuCl$_3$\cite{jasears}, the effect of different antiferromagnetism within the RuCl$_3$ layer on structural and electronics properties is insignificant in determining crystal structures. Therefore we employed a N\'{e}el-type antiferromagnetism in our structural optimizaions. 

\begin{table}
  \centering
  \setlength\extrarowheight{2pt}
  \begin{tabular}{lrr}
    (in \AA) & $\vert {\bf a} \vert $ & $\vert {\bf c} \vert $ \\ \hline
   $\alpha$-RuCl$_3$ & 6.043 & 17.601 \\
   $\alpha$-RuBr$_3$ & 6.404 & 18.634 \\
   $\alpha$-RuI$_3$ & 6.888 & 20.236 
  \end{tabular}
  \caption{
  Lattice parameters of $\alpha$-RuCl$_3$, $\alpha$-RuBr$_3$, and $\alpha$-RuI$_3$ from PBEsol+SCAN optimizations. 
  }
  \label{tab:ac}
\end{table}

Table~\ref{tab:ac} presents PBEsol+SCAN-optimized lattice parameters of all three compounds in the presence of SOC. Our result predicts lattice parameters of $
\alpha$-RuCl$_3$ and $\alpha$-RuI$_3$ to be about 1\% larger than experimentally reported values\cite{Johnson-arXiv,Zhang2021}, but the tendency toward cell expansion as the ligand ion becomes heavier is well-captured within our results.

\begin{figure*}
  \centering
  \includegraphics[width=0.95 \textwidth]{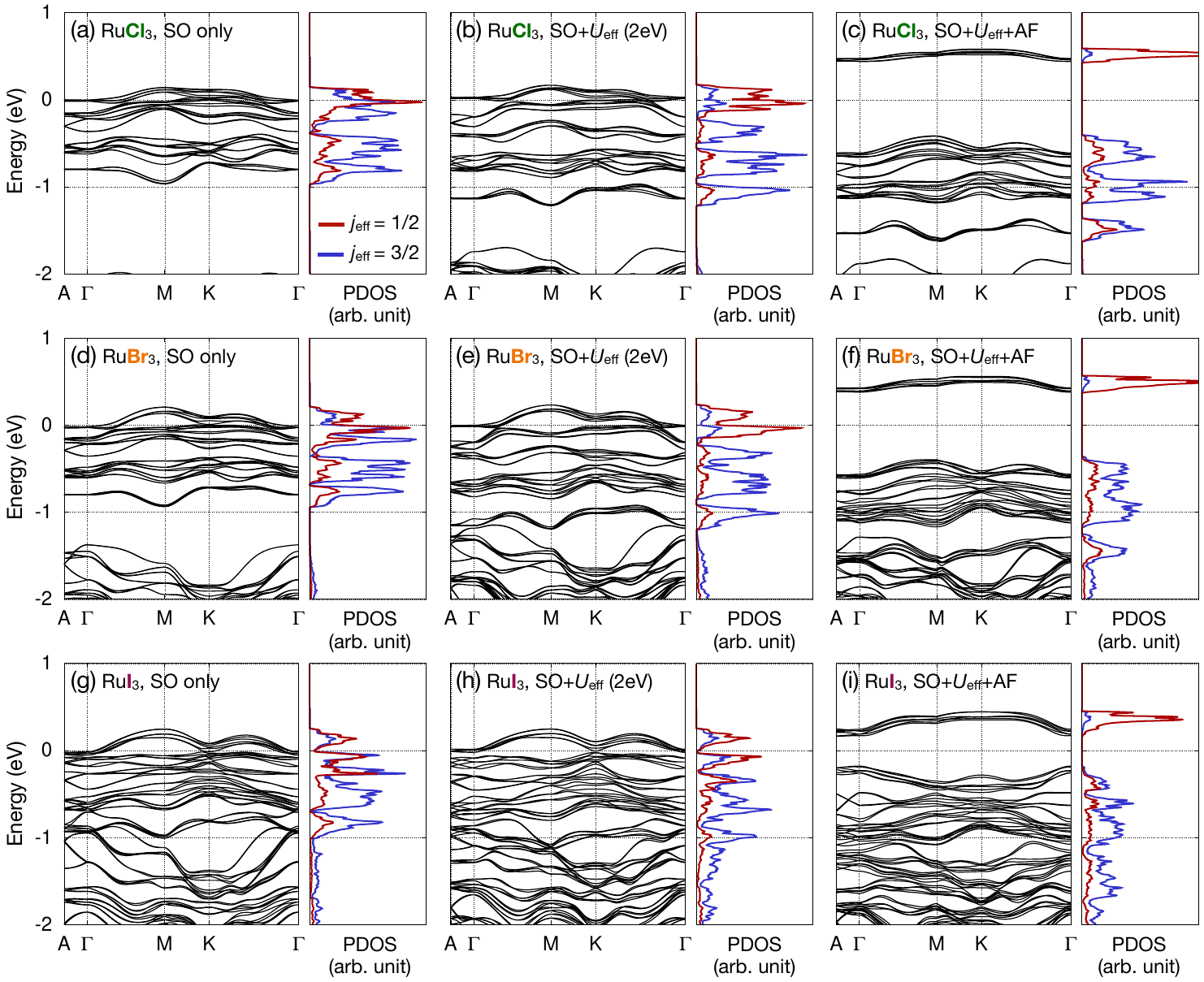}
  \caption{(Color online)
  Band structures and projected densities of states (PDOS) onto the Ru $j_{\rm eff} = 1/2$ and $3/2$ states. Top (a-c), middle (d-f), and bottom (g-i) rows show results from $\alpha$-RuCl$_3$, $\alpha$-RuBr$_3$, and $\alpha$-RuI$_3$, respectively. Left (a,d,g), middle (b,e,h), and right (c,f,i) columns show results with SOC, SOC+$U_{\rm eff}$ = 2 eV, and SOC+$U_{\rm eff}$ = 2 eV in the presence of the N\'{e}el-type antiferromagnetism, respectively. Note that nonmagnetic constraint was enforced when obtaining results in the left and middle columns. 
  }
  \label{fig:bandndos}
\end{figure*}

\section{Electronic structure and the spin-orbital-entangled states}

Fig.~\ref{fig:bandndos} summarizes our electronic structure calculation results for all three compounds, showing the band structures and projected densities of states (PDOS) onto the spin-orbit-entangled $j_{\rm eff} = 1/2$ and $3/2$ states. Note that the Ru $j_{\rm eff}$ states, eigenstates of the SOC hamiltonian within the Ru $t_{\rm 2g}$ orbital, are as follows,

\begin{align*}
\left\vert j_{\rm eff} = \frac{1}{2};\pm \frac{1}{2} \right\rangle &= 
\mp \frac{1}{\sqrt{3}} (\vert d_{xy}, \uparrow \downarrow \rangle \pm \vert d_{yz}, \downarrow\uparrow \rangle + i \vert d_{xz}, \downarrow \uparrow \rangle) \\
\left\vert j_{\rm eff} = \frac{3}{2};\pm \frac{1}{2} \right\rangle &= 
 \sqrt{\frac{2}{3}} \left(\vert d_{xy}, \uparrow \downarrow \rangle 
 \mp \frac{
 	\vert d_{yz}, \downarrow \uparrow \rangle \pm i \vert d_{xz}, \downarrow \uparrow \rangle	
	}{2}
  \right) \\
  \left\vert j_{\rm eff} = \frac{3}{2};\pm \frac{3}{2} \right\rangle &= 
  \mp \frac{1}{\sqrt{2}} ( 
  \vert d_{yz}, \uparrow\downarrow \rangle \pm i \vert d_{xz}, \uparrow \downarrow \rangle.
  )
\end{align*}

Comparing the left, center, and right columns in Fig.~\ref{fig:bandndos}, it can be seen that the inclusion of the on-site Coulomb interaction enhances the splitting between the $j_{\rm eff} = 1/2$ and $3/2$ states, namely the effective magnitude of SOC. Note that this effect, the enhancement of SOC by $U_{\rm eff}$, has been discussed in previous works on $\alpha$-RuCl$_3$\cite{hskim_RuCl3} (and also in Ref.~\onlinecite{Zhang2021}). Inclusion of magnetism (and the resulting opening the band gap) enhances the splitting even further (compare the middle and right columns in Fig.~\ref{fig:bandndos}), therefore making $\alpha$-RuCl$_3$ and $\alpha$-RuI$_3$ good Kitaev magnet candidates. Note that $\alpha$-RuI$_3$ has been reported to be metallic\cite{Ni2021}, contrary to our result shown in Fig.~\ref{fig:bandndos}(i). The discrepancy may originate from technical details (for example, the choice of local atomic projectors employed in the DFT+$U_{\rm eff}$ methodology). We'd like to point out that in our results the band gap size in $\alpha$-RuI$_3$ is much smaller than that of $\alpha$-RuCl$_3$ and $\alpha$-RuBr$_3$ (compare Fig.~\ref{fig:bandndos}(c), (f), and (i)), consistent with the fact that larger $d$-$p$ hybridization in $\alpha$-RuI$_3$ induces larger bandwidth and weaker electron correlation effects. Additionally, the enhancement of the $j_{\rm eff} = 1/2$-$3/2$ splitting introduced by $U_{\rm eff}$ is not significant (Fig.~\ref{fig:bandndos}(g) vs (h)) compared to other compounds, possibly due to the larger $d$-$p$ hybridization in $\alpha$-RuI$_3$.

\begin{figure}
  \centering
  \includegraphics[width=0.45 \textwidth]{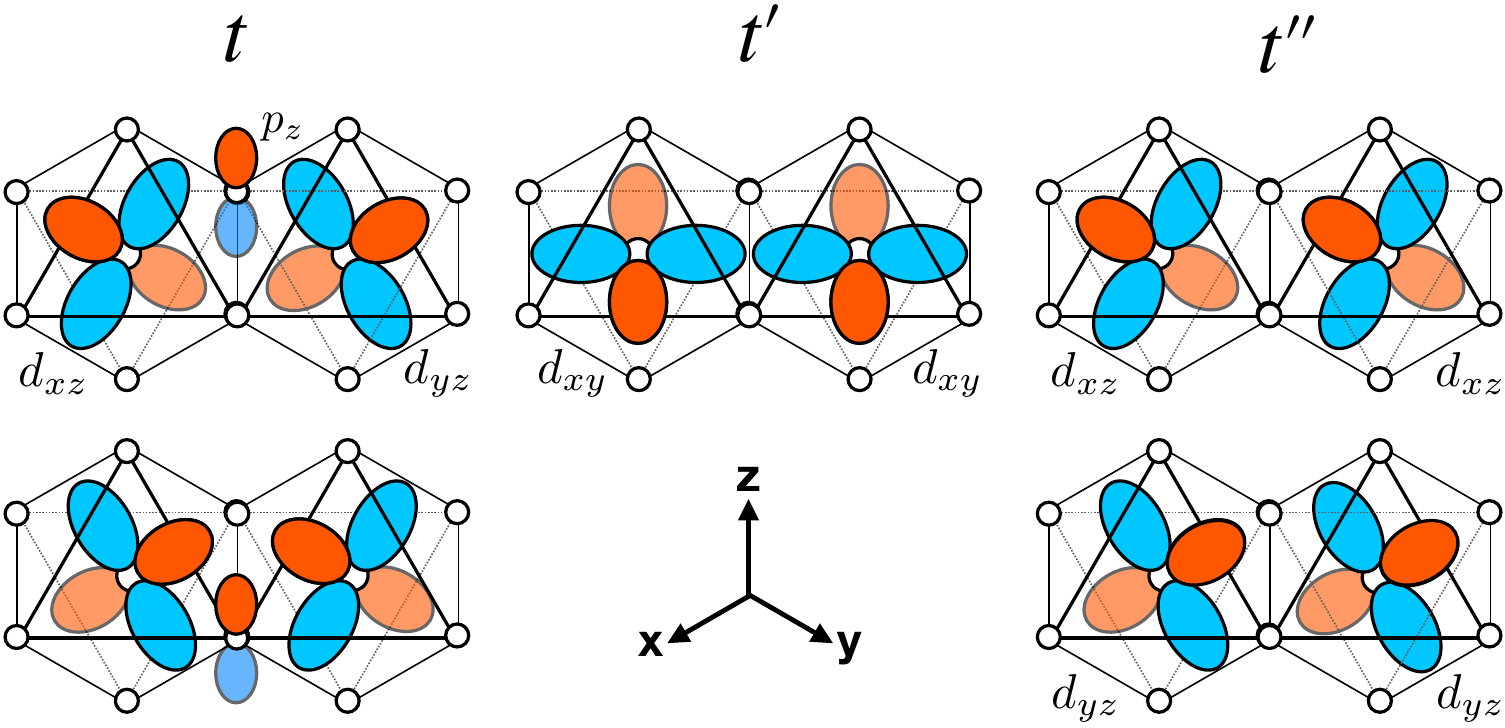}
  \caption{(Color online)
  Illustration of major hopping channels between nearest-neighboring Ru $t_{\rm 2g}$ orbitals.  
  }
  \label{fig:hopping}
\end{figure}

\section{Electron hopping channels and magnetic exchange interactions}

Having established the spin-orbit-coupled nature of our systems, we next switch to magnetic exchange interactions. Fig.~\ref{fig:hopping} represents major nearest-neighbor hopping channels between Ru $t_{\rm 2g}$ orbitals. Among those the first $t$-term (left panels in Fig.~\ref{fig:hopping}), which is mediated via an intermediate anion $p$-orbital, is critical to induce the Kitaev interaction, while the others tend to give rise to conventional Heisenberg interactions or symmetric anisotropy interactions. Specifically, the general magnetic Hamiltonian in $\alpha$-RuX$_3$ has the following approximate form\cite{rau2014generic},

\begin{align*}
  H &\simeq \sum_{\mathclap{\langle ij\rangle\in \alpha\beta(\gamma)}} \left(J {\bf S}_{i}\cdot{\bf S}_{j} + K S^\gamma_i S^\gamma_j + \Gamma(S^\alpha_iS^\beta_j + S^\beta_i S^\alpha_j) \right) \nonumber \\
    &+ \sum_{\mathclap{\langle \langle \langle ij\rangle \rangle \rangle \in \alpha\beta(\gamma)}} \left(J_3 {\bf S}_i \cdot {\bf S}_j )\right), \nonumber
\end{align*}
where $\alpha, \beta, \gamma$ denote bond directions and the relevant spin components ($x,y,z$) involved in the Kitaev interactions. Note that $\langle ij \rangle$ and $\langle \langle \langle ij \rangle \rangle \rangle$ mean nearest-neighbor and third-nearest-neighbor $i$-$j$ sites, respectively. The $J$, $K$, and $\Gamma$ expressed in terms of hopping intergrals and Coulomb interaction parameters are explicitly given via perturbation theory as follows\cite{rau2014generic},

\begin{align*}
J &= \frac{4}{27}\left[ 
\frac{6t'' (t''+2t')}{U-3J_{\rm H}} + 
\frac{2(t''-t')^2}{U-J_{\rm H}} + \frac{(2t'' + t')^2}{U+2J_{\rm H}} 
\right] \nonumber \\
K &= \frac{8J_{\rm H}}{9}\left[ 
\frac{(t''-t')^2 - 3t^2}{(U-3J_{\rm H})(U-J_{\rm H})} 
\right] \nonumber \\
\Gamma &= \frac{16J_{\rm H}}{9}\left[ 
\frac{t(t''-t')}{(U-3J_{\rm H})(U-J_{\rm H})}
\right]. 
\end{align*}

Here $U$ and $J_{\rm H}$ are the strength of the on-site Coulomb repulsion (not necessarily same with $U_{\rm eff}$) and Hund's coupling within the Ru $t_{\rm 2g}$ orbital. 

\begin{table}
  \centering
  \setlength\extrarowheight{2pt}
  \begin{tabular}{lrrrrr}
    (in eV) & $t$ & $t'$ & $t''$ & $t_{\rm 3rd}$ & $t^{\rm max}_{\rm inter}$ \\ \hline
   $\alpha$-RuCl$_3$ & +0.184 & -0.054 & +0.035 & -0.041 & -0.028 \\
   $\alpha$-RuBr$_3$ & +0.169 & -0.030 & +0.024 & -0.040 & -0.040 \\
   $\alpha$-RuI$_3$   & +0.170 & +0.007 & +0.009 & -0.050 & -0.049
  \end{tabular}
  \caption{
  Major nearest-neighbor hopping integrals ($t,t',t''$) and largest third-nearest-neighbor ($t_{\rm 3rd}$) and interlayer ($t^{\rm max}_{\rm inter}$) hopping terms from Wannier orbital method. 
  }
  \label{tab:hop}
\end{table}

The hopping terms $t$, $t'$, and $t''$ can be obtained via employing Wannier function method\cite{MLWF1,MLWF2} whose magnitudes were tabulated in Table~\ref{tab:hop}. Interestingly, the magnitude of the $d$-$p$-$d$ hopping term $t$ decreases slightly as we move from $\alpha$-RuCl$_3$ to $\alpha$-RuI$_3$. This is surprising because larger $d$-$p$-hybridizations between Ru $d$- and anion $p$-orbitals, which directly contribute to the $t$ channel, are expected to be enhanced as the anion becomes heavier. The origin of this behavior is unclear, and it can be speculated that the presence of multiple $d$-$d$ and $d$-$p$-$d$ hopping channels may induce slight reduction of $t$ in RuBr$_3$. On the other hand, the magnitude of $t'$ and $t''$, which introduce Heisenberg $J$, decreases when Cl is replaced by Br or I. This observation is easier to understand because larger anion size makes the nearest Ru-Ru distance farther, resulting in suppressed direct $d$-$d$ overlap channels like $t'$ and $t''$.

\begin{figure}
  \centering
  \includegraphics[width=0.45 \textwidth]{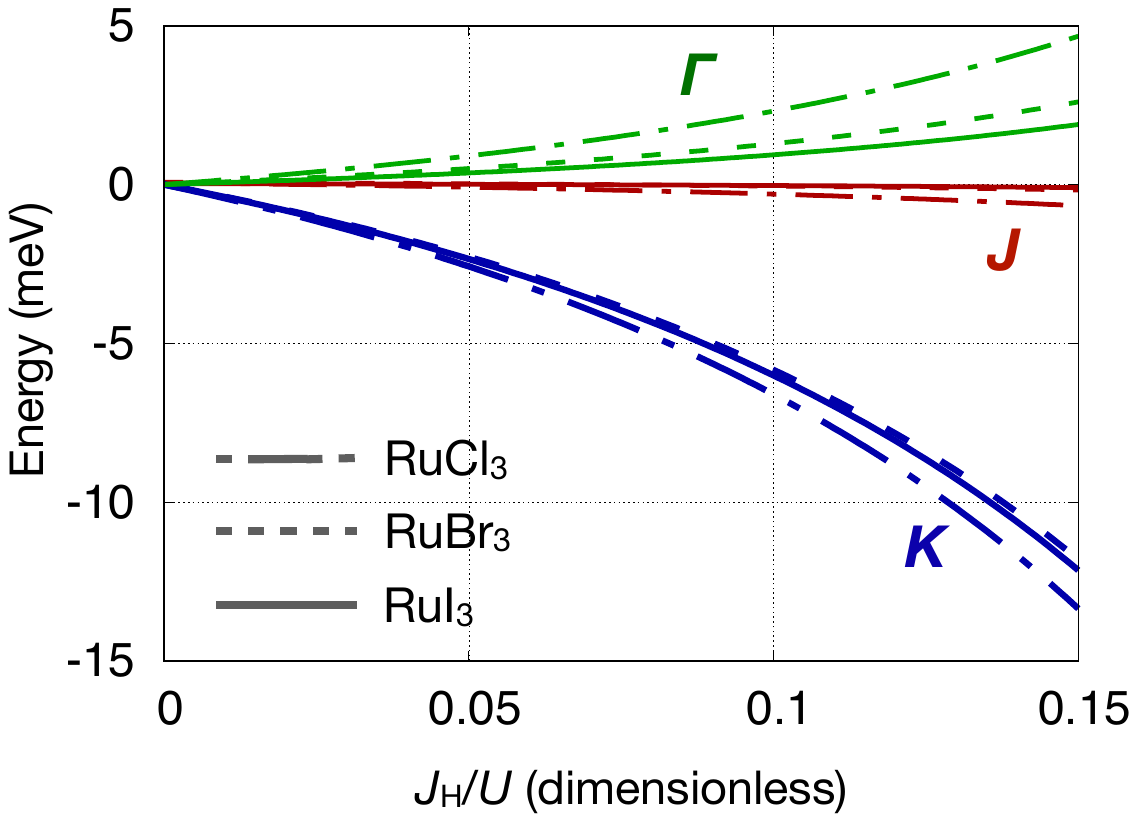}
  \caption{(Color online)
  Nearest-neighbor magnetic exchange interactions for $\alpha$-RuCl$_3$, $\alpha$-RuBr$_3$, and $\alpha$-RuI$_3$ as a function of $J_{\rm H} / U$. 
  }
  \label{fig:ex}
\end{figure}

Because we have hopping parameters from {\it ab-initio} calculations, now we can estimate the magnitudes of $J$, $K$, and $\Gamma$ as a function of $U$ and $J_{\rm H}$. We set $U = U_{\rm eff} = $ 2 eV, and compute exchange interactions as a function of $J_{\rm H} / U$ (note that different values of $U$ just changes the overall scale of the exchange parameters). Fig.~\ref{fig:ex} shows the results, and thanks to the diminishing $t'$ and $t''$ in $\alpha$-RuBr$_3$ and $\alpha$-RuI$_3$, the size of $J$ and $\Gamma$ reduces as we move from $\alpha$-RuCl$_3$ to $\alpha$-RuI$_3$. Since $\alpha$-RuI$_3$ has been reported to be metal, $\alpha$-RuBr$_3$ seems to be the best candidate to host Kitaev magnetism in this series of compounds.

We conclude this section by mentioning third-nearest-neighbor and inter-layer hopping elements. Table~\ref{tab:hop} shows that both terms increase as the ligand anion becomes heavier, as expected in the beginning. The enhancement is, however, not that significant compared to the changes in $t'$ and $t''$. Although many different channels can constructively contribute to further neighbor and interlayer hopping channels, as discussed in Ref.~\onlinecite{Winter2016,Catuneanu2016}, we may speculate that the enhancement in further neighbor and interlayer hopping channels may not that substantial in $\alpha$-RuBr$_3$ and $\alpha$-RuI$_3$, which makes $\alpha$-RuBr$_3$ a promising candidate to realize Kitaev magnetism in this family. 

\section{summary}
In this work we performed a preliminary first-principles density functional theory calculations for $\alpha$-RuCl$_3$, $\alpha$-RuBr$_3$, and $\alpha$-RuI$_3$, making an assessment of their viability to realize Kitaev magnetism. From our results it can be seen that $\alpha$-RuCl$_3$ and $\alpha$-RuBr$_3$ may host the spin-orbit-entangled $j_{\rm eff} = 1/2$ moment, and that $\alpha$-RuBr$_3$ shows even stronger Kitaev magnetism compared to $\alpha$-RuCl$_3$. 

Note that some possible effects of minor lattice distortions, for example the effect of trigonal crystal fields, were not discussed here\cite{rau_trigonal}. Estimation of next- and third-nearest-neighbor Heisenberg interactions that induce long-range magnetic orders needs to be done as well for further studies on magnetic properties of these systems. Hence continuing theoretical and experimental studies are needed at this moment, but we believe that $\alpha$-RuBr$_3$ can be another interesting system to study Kitaev physics in addition to $\alpha$-RuCl$_3$. One last thing to comment is, since $\alpha$-RuI$_3$ has been reported to be nonmagnetic and metallic, it has potentials to host topological band insulating phases with weak-to-intermeadiate electron correlations as previously suggested\cite{Catuneanu2016}.

\begin{acknowledgments}
We thank K.-Y. Choi for useful discussions, and acknowledge support from the Korea Research Fellow (KRF) Program and the Basic Science Research Program through the National Research Foundation of Korea funded by the Ministry of Education (NRF-2019H1D3A1A01102984 and NRF-2020R1C1C1005900). We also thank the support of computational resources including technical assistance from the National Supercomputing Center of Korea (Grant No. KSC-2020-CRE-0156) 
\end{acknowledgments}


\bibliography{rux3}

\begin{thebibliography}{37}%
\makeatletter
\providecommand \@ifxundefined [1]{%
 \@ifx{#1\undefined}
}%
\providecommand \@ifnum [1]{%
 \ifnum #1\expandafter \@firstoftwo
 \else \expandafter \@secondoftwo
 \fi
}%
\providecommand \@ifx [1]{%
 \ifx #1\expandafter \@firstoftwo
 \else \expandafter \@secondoftwo
 \fi
}%
\providecommand \natexlab [1]{#1}%
\providecommand \enquote  [1]{``#1''}%
\providecommand \bibnamefont  [1]{#1}%
\providecommand \bibfnamefont [1]{#1}%
\providecommand \citenamefont [1]{#1}%
\providecommand \href@noop [0]{\@secondoftwo}%
\providecommand \href [0]{\begingroup \@sanitize@url \@href}%
\providecommand \@href[1]{\@@startlink{#1}\@@href}%
\providecommand \@@href[1]{\endgroup#1\@@endlink}%
\providecommand \@sanitize@url [0]{\catcode `\\12\catcode `\$12\catcode
  `\&12\catcode `\#12\catcode `\^12\catcode `\_12\catcode `\%12\relax}%
\providecommand \@@startlink[1]{}%
\providecommand \@@endlink[0]{}%
\providecommand \url  [0]{\begingroup\@sanitize@url \@url }%
\providecommand \@url [1]{\endgroup\@href {#1}{\urlprefix }}%
\providecommand \urlprefix  [0]{URL }%
\providecommand \Eprint [0]{\href }%
\providecommand \doibase [0]{http://dx.doi.org/}%
\providecommand \selectlanguage [0]{\@gobble}%
\providecommand \bibinfo  [0]{\@secondoftwo}%
\providecommand \bibfield  [0]{\@secondoftwo}%
\providecommand \translation [1]{[#1]}%
\providecommand \BibitemOpen [0]{}%
\providecommand \bibitemStop [0]{}%
\providecommand \bibitemNoStop [0]{.\EOS\space}%
\providecommand \EOS [0]{\spacefactor3000\relax}%
\providecommand \BibitemShut  [1]{\csname bibitem#1\endcsname}%
\let\auto@bib@innerbib\@empty
\bibitem [{\citenamefont {Kitaev}(2006)}]{kitaev2006anyons}%
  \BibitemOpen
  \bibfield  {author} {\bibinfo {author} {\bibfnamefont {A.}~\bibnamefont
  {Kitaev}},\ }\href
  {http://www.sciencedirect.com/science/article/pii/S0003491605002381}
  {\bibfield  {journal} {\bibinfo  {journal} {Ann. Phys.}\ }\textbf {\bibinfo
  {volume} {321}},\ \bibinfo {pages} {2} (\bibinfo {year} {2006})}\BibitemShut
  {NoStop}%
\bibitem [{\citenamefont {Jackeli}\ and\ \citenamefont
  {Khaliullin}(2009)}]{jackeli2009mott}%
  \BibitemOpen
  \bibfield  {author} {\bibinfo {author} {\bibfnamefont {G.}~\bibnamefont
  {Jackeli}}\ and\ \bibinfo {author} {\bibfnamefont {G.}~\bibnamefont
  {Khaliullin}},\ }\href
  {http://link.aps.org/doi/10.1103/PhysRevLett.102.017205} {\bibfield
  {journal} {\bibinfo  {journal} {Phys. Rev. Lett}\ }\textbf {\bibinfo {volume}
  {102}},\ \bibinfo {pages} {017205} (\bibinfo {year} {2009})}\BibitemShut
  {NoStop}%
\bibitem [{\citenamefont {Winter}\ \emph {et~al.}(2017)\citenamefont {Winter},
  \citenamefont {Tsirlin}, \citenamefont {Daghofer}, \citenamefont {van~den
  Brink}, \citenamefont {Singh}, \citenamefont {Gegenwart},\ and\ \citenamefont
  {Valent{\'{\i}}}}]{Winter2017}%
  \BibitemOpen
  \bibfield  {author} {\bibinfo {author} {\bibfnamefont {S.~M.}\ \bibnamefont
  {Winter}}, \bibinfo {author} {\bibfnamefont {A.~A.}\ \bibnamefont {Tsirlin}},
  \bibinfo {author} {\bibfnamefont {M.}~\bibnamefont {Daghofer}}, \bibinfo
  {author} {\bibfnamefont {J.}~\bibnamefont {van~den Brink}}, \bibinfo {author}
  {\bibfnamefont {Y.}~\bibnamefont {Singh}}, \bibinfo {author} {\bibfnamefont
  {P.}~\bibnamefont {Gegenwart}}, \ and\ \bibinfo {author} {\bibfnamefont
  {R.}~\bibnamefont {Valent{\'{\i}}}},\ }\href {\doibase
  10.1088/1361-648x/aa8cf5} {\bibfield  {journal} {\bibinfo  {journal} {Journal
  of Physics: Condensed Matter}\ }\textbf {\bibinfo {volume} {29}},\ \bibinfo
  {pages} {493002} (\bibinfo {year} {2017})}\BibitemShut {NoStop}%
\bibitem [{\citenamefont {Takagi}\ \emph {et~al.}(2019)\citenamefont {Takagi},
  \citenamefont {Takayama}, \citenamefont {Jackeli}, \citenamefont
  {Khaliullin},\ and\ \citenamefont {Nagler}}]{KitaevReview}%
  \BibitemOpen
  \bibfield  {author} {\bibinfo {author} {\bibfnamefont {H.}~\bibnamefont
  {Takagi}}, \bibinfo {author} {\bibfnamefont {T.}~\bibnamefont {Takayama}},
  \bibinfo {author} {\bibfnamefont {G.}~\bibnamefont {Jackeli}}, \bibinfo
  {author} {\bibfnamefont {G.}~\bibnamefont {Khaliullin}}, \ and\ \bibinfo
  {author} {\bibfnamefont {S.~E.}\ \bibnamefont {Nagler}},\ }\href {\doibase
  10.1038/s42254-019-0038-2} {\bibfield  {journal} {\bibinfo  {journal} {Nature
  Reviews Physics}\ }\textbf {\bibinfo {volume} {1}},\ \bibinfo {pages} {264}
  (\bibinfo {year} {2019})}\BibitemShut {NoStop}%
\bibitem [{\citenamefont {Plumb}\ \emph {et~al.}(2014)\citenamefont {Plumb},
  \citenamefont {Clancy}, \citenamefont {Sandilands}, \citenamefont
  {Vijay~Shankar}, \citenamefont {Hu}, \citenamefont {Burch}, \citenamefont
  {Kee},\ and\ \citenamefont {Kim}}]{plumb2014alpha}%
  \BibitemOpen
  \bibfield  {author} {\bibinfo {author} {\bibfnamefont {K.~W.}\ \bibnamefont
  {Plumb}}, \bibinfo {author} {\bibfnamefont {J.~P.}\ \bibnamefont {Clancy}},
  \bibinfo {author} {\bibfnamefont {L.~J.}\ \bibnamefont {Sandilands}},
  \bibinfo {author} {\bibfnamefont {V.}~\bibnamefont {Vijay~Shankar}}, \bibinfo
  {author} {\bibfnamefont {Y.~F.}\ \bibnamefont {Hu}}, \bibinfo {author}
  {\bibfnamefont {K.~S.}\ \bibnamefont {Burch}}, \bibinfo {author}
  {\bibfnamefont {H.-Y.}\ \bibnamefont {Kee}}, \ and\ \bibinfo {author}
  {\bibfnamefont {Y.-J.}\ \bibnamefont {Kim}},\ }\href {\doibase
  10.1103/PhysRevB.90.041112} {\bibfield  {journal} {\bibinfo  {journal} {Phys.
  Rev. B}\ }\textbf {\bibinfo {volume} {90}},\ \bibinfo {pages} {041112}
  (\bibinfo {year} {2014})}\BibitemShut {NoStop}%
\bibitem [{\citenamefont {Kim}\ \emph {et~al.}(2015)\citenamefont {Kim},
  \citenamefont {V.}, \citenamefont {Catuneanu},\ and\ \citenamefont
  {Kee}}]{hskim_RuCl3}%
  \BibitemOpen
  \bibfield  {author} {\bibinfo {author} {\bibfnamefont {H.-S.}\ \bibnamefont
  {Kim}}, \bibinfo {author} {\bibfnamefont {V.~S.}\ \bibnamefont {V.}},
  \bibinfo {author} {\bibfnamefont {A.}~\bibnamefont {Catuneanu}}, \ and\
  \bibinfo {author} {\bibfnamefont {H.-Y.}\ \bibnamefont {Kee}},\ }\href
  {\doibase 10.1103/PhysRevB.91.241110} {\bibfield  {journal} {\bibinfo
  {journal} {Phys. Rev. B}\ }\textbf {\bibinfo {volume} {91}},\ \bibinfo
  {pages} {241110} (\bibinfo {year} {2015})}\BibitemShut {NoStop}%
\bibitem [{\citenamefont {Sears}\ \emph {et~al.}(2015)\citenamefont {Sears},
  \citenamefont {Songvilay}, \citenamefont {Plumb}, \citenamefont {Clancy},
  \citenamefont {Qiu}, \citenamefont {Zhao}, \citenamefont {Parshall},\ and\
  \citenamefont {Kim}}]{jasears}%
  \BibitemOpen
  \bibfield  {author} {\bibinfo {author} {\bibfnamefont {J.~A.}\ \bibnamefont
  {Sears}}, \bibinfo {author} {\bibfnamefont {M.}~\bibnamefont {Songvilay}},
  \bibinfo {author} {\bibfnamefont {K.~W.}\ \bibnamefont {Plumb}}, \bibinfo
  {author} {\bibfnamefont {J.~P.}\ \bibnamefont {Clancy}}, \bibinfo {author}
  {\bibfnamefont {Y.}~\bibnamefont {Qiu}}, \bibinfo {author} {\bibfnamefont
  {Y.}~\bibnamefont {Zhao}}, \bibinfo {author} {\bibfnamefont {D.}~\bibnamefont
  {Parshall}}, \ and\ \bibinfo {author} {\bibfnamefont {Y.-J.}\ \bibnamefont
  {Kim}},\ }\href {\doibase 10.1103/PhysRevB.91.144420} {\bibfield  {journal}
  {\bibinfo  {journal} {Phys. Rev. B}\ }\textbf {\bibinfo {volume} {91}},\
  \bibinfo {pages} {144420} (\bibinfo {year} {2015})}\BibitemShut {NoStop}%
\bibitem [{\citenamefont {Sandilands}\ \emph {et~al.}(2015)\citenamefont
  {Sandilands}, \citenamefont {Tian}, \citenamefont {Plumb}, \citenamefont
  {Kim},\ and\ \citenamefont {Burch}}]{luke2}%
  \BibitemOpen
  \bibfield  {author} {\bibinfo {author} {\bibfnamefont {L.~J.}\ \bibnamefont
  {Sandilands}}, \bibinfo {author} {\bibfnamefont {Y.}~\bibnamefont {Tian}},
  \bibinfo {author} {\bibfnamefont {K.~W.}\ \bibnamefont {Plumb}}, \bibinfo
  {author} {\bibfnamefont {Y.-J.}\ \bibnamefont {Kim}}, \ and\ \bibinfo
  {author} {\bibfnamefont {K.~S.}\ \bibnamefont {Burch}},\ }\href {\doibase
  10.1103/PhysRevLett.114.147201} {\bibfield  {journal} {\bibinfo  {journal}
  {Phys. Rev. Lett.}\ }\textbf {\bibinfo {volume} {114}},\ \bibinfo {pages}
  {147201} (\bibinfo {year} {2015})}\BibitemShut {NoStop}%
\bibitem [{\citenamefont {Sandilands}\ \emph {et~al.}(2016)\citenamefont
  {Sandilands}, \citenamefont {Tian}, \citenamefont {Reijnders}, \citenamefont
  {Kim}, \citenamefont {Plumb}, \citenamefont {Kim}, \citenamefont {Kee},\ and\
  \citenamefont {Burch}}]{luke}%
  \BibitemOpen
  \bibfield  {author} {\bibinfo {author} {\bibfnamefont {L.~J.}\ \bibnamefont
  {Sandilands}}, \bibinfo {author} {\bibfnamefont {Y.}~\bibnamefont {Tian}},
  \bibinfo {author} {\bibfnamefont {A.~A.}\ \bibnamefont {Reijnders}}, \bibinfo
  {author} {\bibfnamefont {H.-S.}\ \bibnamefont {Kim}}, \bibinfo {author}
  {\bibfnamefont {K.~W.}\ \bibnamefont {Plumb}}, \bibinfo {author}
  {\bibfnamefont {Y.-J.}\ \bibnamefont {Kim}}, \bibinfo {author} {\bibfnamefont
  {H.-Y.}\ \bibnamefont {Kee}}, \ and\ \bibinfo {author} {\bibfnamefont
  {K.~S.}\ \bibnamefont {Burch}},\ }\href {\doibase 10.1103/PhysRevB.93.075144}
  {\bibfield  {journal} {\bibinfo  {journal} {Phys. Rev. B}\ }\textbf {\bibinfo
  {volume} {93}},\ \bibinfo {pages} {075144} (\bibinfo {year}
  {2016})}\BibitemShut {NoStop}%
\bibitem [{\citenamefont {Kim}\ and\ \citenamefont {Kee}(2016)}]{HSK2016}%
  \BibitemOpen
  \bibfield  {author} {\bibinfo {author} {\bibfnamefont {H.-S.}\ \bibnamefont
  {Kim}}\ and\ \bibinfo {author} {\bibfnamefont {H.-Y.}\ \bibnamefont {Kee}},\
  }\href {\doibase 10.1103/PhysRevB.93.155143} {\bibfield  {journal} {\bibinfo
  {journal} {Phys. Rev. B}\ }\textbf {\bibinfo {volume} {93}},\ \bibinfo
  {pages} {155143} (\bibinfo {year} {2016})}\BibitemShut {NoStop}%
\bibitem [{\citenamefont {Banerjee}\ \emph {et~al.}(2015)\citenamefont
  {Banerjee}, \citenamefont {Bridges}, \citenamefont {Yan}, \citenamefont
  {Aczel}, \citenamefont {Li}, \citenamefont {Stone}, \citenamefont {Granroth},
  \citenamefont {Lumsden}, \citenamefont {Yiu}, \citenamefont {Knolle},
  \citenamefont {Kovrizhin}, \citenamefont {Bhattacharjee}, \citenamefont
  {Moessner}, \citenamefont {Tennant}, \citenamefont {Mandrus},\ and\
  \citenamefont {Nagler}}]{Banerjee}%
  \BibitemOpen
  \bibfield  {author} {\bibinfo {author} {\bibfnamefont {A.}~\bibnamefont
  {Banerjee}}, \bibinfo {author} {\bibfnamefont {C.}~\bibnamefont {Bridges}},
  \bibinfo {author} {\bibfnamefont {J.-Q.}\ \bibnamefont {Yan}}, \bibinfo
  {author} {\bibfnamefont {A.}~\bibnamefont {Aczel}}, \bibinfo {author}
  {\bibfnamefont {L.}~\bibnamefont {Li}}, \bibinfo {author} {\bibfnamefont
  {M.}~\bibnamefont {Stone}}, \bibinfo {author} {\bibfnamefont
  {G.}~\bibnamefont {Granroth}}, \bibinfo {author} {\bibfnamefont
  {M.}~\bibnamefont {Lumsden}}, \bibinfo {author} {\bibfnamefont
  {Y.}~\bibnamefont {Yiu}}, \bibinfo {author} {\bibfnamefont {J.}~\bibnamefont
  {Knolle}}, \bibinfo {author} {\bibfnamefont {D.}~\bibnamefont {Kovrizhin}},
  \bibinfo {author} {\bibnamefont {Bhattacharjee}}, \bibinfo {author}
  {\bibfnamefont {R.}~\bibnamefont {Moessner}}, \bibinfo {author}
  {\bibfnamefont {D.}~\bibnamefont {Tennant}}, \bibinfo {author} {\bibfnamefont
  {D.}~\bibnamefont {Mandrus}}, \ and\ \bibinfo {author} {\bibfnamefont
  {S.}~\bibnamefont {Nagler}},\ }\href {http://arxiv.org/abs/1504.08037}
  {\bibfield  {journal} {\bibinfo  {journal} {arXiv preprint arXiv:1504.08037}\
  } (\bibinfo {year} {2015})}\BibitemShut {NoStop}%
\bibitem [{\citenamefont {Do}\ \emph {et~al.}(2017)\citenamefont {Do},
  \citenamefont {Park}, \citenamefont {Yoshitake}, \citenamefont {Nasu},
  \citenamefont {Motome}, \citenamefont {Kwon}, \citenamefont {Adroja},
  \citenamefont {Voneshen}, \citenamefont {Kim}, \citenamefont {Jang},
  \citenamefont {Park}, \citenamefont {Choi},\ and\ \citenamefont
  {Ji}}]{Do2017}%
  \BibitemOpen
  \bibfield  {author} {\bibinfo {author} {\bibfnamefont {S.-H.}\ \bibnamefont
  {Do}}, \bibinfo {author} {\bibfnamefont {S.-Y.}\ \bibnamefont {Park}},
  \bibinfo {author} {\bibfnamefont {J.}~\bibnamefont {Yoshitake}}, \bibinfo
  {author} {\bibfnamefont {J.}~\bibnamefont {Nasu}}, \bibinfo {author}
  {\bibfnamefont {Y.}~\bibnamefont {Motome}}, \bibinfo {author} {\bibfnamefont
  {Y.~S.}\ \bibnamefont {Kwon}}, \bibinfo {author} {\bibfnamefont {D.~T.}\
  \bibnamefont {Adroja}}, \bibinfo {author} {\bibfnamefont {D.~J.}\
  \bibnamefont {Voneshen}}, \bibinfo {author} {\bibfnamefont {K.}~\bibnamefont
  {Kim}}, \bibinfo {author} {\bibfnamefont {T.~H.}\ \bibnamefont {Jang}},
  \bibinfo {author} {\bibfnamefont {J.~H.}\ \bibnamefont {Park}}, \bibinfo
  {author} {\bibfnamefont {K.-Y.}\ \bibnamefont {Choi}}, \ and\ \bibinfo
  {author} {\bibfnamefont {S.}~\bibnamefont {Ji}},\ }\href {\doibase
  10.1038/nphys4264} {\bibfield  {journal} {\bibinfo  {journal} {Nature
  Physics}\ }\textbf {\bibinfo {volume} {13}},\ \bibinfo {pages} {1079}
  (\bibinfo {year} {2017})}\BibitemShut {NoStop}%
\bibitem [{\citenamefont {Kasahara}\ \emph {et~al.}(2018)\citenamefont
  {Kasahara}, \citenamefont {Ohnishi}, \citenamefont {Mizukami}, \citenamefont
  {Tanaka}, \citenamefont {Ma}, \citenamefont {Sugii}, \citenamefont {Kurita},
  \citenamefont {Tanaka}, \citenamefont {Nasu}, \citenamefont {Motome},
  \citenamefont {Shibauchi},\ and\ \citenamefont {Matsuda}}]{Kasahara2018}%
  \BibitemOpen
  \bibfield  {author} {\bibinfo {author} {\bibfnamefont {Y.}~\bibnamefont
  {Kasahara}}, \bibinfo {author} {\bibfnamefont {T.}~\bibnamefont {Ohnishi}},
  \bibinfo {author} {\bibfnamefont {Y.}~\bibnamefont {Mizukami}}, \bibinfo
  {author} {\bibfnamefont {O.}~\bibnamefont {Tanaka}}, \bibinfo {author}
  {\bibfnamefont {S.}~\bibnamefont {Ma}}, \bibinfo {author} {\bibfnamefont
  {K.}~\bibnamefont {Sugii}}, \bibinfo {author} {\bibfnamefont
  {N.}~\bibnamefont {Kurita}}, \bibinfo {author} {\bibfnamefont
  {H.}~\bibnamefont {Tanaka}}, \bibinfo {author} {\bibfnamefont
  {J.}~\bibnamefont {Nasu}}, \bibinfo {author} {\bibfnamefont {Y.}~\bibnamefont
  {Motome}}, \bibinfo {author} {\bibfnamefont {T.}~\bibnamefont {Shibauchi}}, \
  and\ \bibinfo {author} {\bibfnamefont {Y.}~\bibnamefont {Matsuda}},\ }\href
  {\doibase 10.1038/s41586-018-0274-0} {\bibfield  {journal} {\bibinfo
  {journal} {Nature}\ }\textbf {\bibinfo {volume} {559}},\ \bibinfo {pages}
  {227} (\bibinfo {year} {2018})}\BibitemShut {NoStop}%
\bibitem [{\citenamefont {Sears}\ \emph {et~al.}(2020)\citenamefont {Sears},
  \citenamefont {Chern}, \citenamefont {Kim}, \citenamefont {Bereciartua},
  \citenamefont {Francoual}, \citenamefont {Kim},\ and\ \citenamefont
  {Kim}}]{Sears2020}%
  \BibitemOpen
  \bibfield  {author} {\bibinfo {author} {\bibfnamefont {J.~A.}\ \bibnamefont
  {Sears}}, \bibinfo {author} {\bibfnamefont {L.~E.}\ \bibnamefont {Chern}},
  \bibinfo {author} {\bibfnamefont {S.}~\bibnamefont {Kim}}, \bibinfo {author}
  {\bibfnamefont {P.~J.}\ \bibnamefont {Bereciartua}}, \bibinfo {author}
  {\bibfnamefont {S.}~\bibnamefont {Francoual}}, \bibinfo {author}
  {\bibfnamefont {Y.~B.}\ \bibnamefont {Kim}}, \ and\ \bibinfo {author}
  {\bibfnamefont {Y.-J.}\ \bibnamefont {Kim}},\ }\href {\doibase
  10.1038/s41567-020-0874-0} {\bibfield  {journal} {\bibinfo  {journal} {Nature
  Physics}\ }\textbf {\bibinfo {volume} {16}},\ \bibinfo {pages} {837}
  (\bibinfo {year} {2020})}\BibitemShut {NoStop}%
\bibitem [{\citenamefont {Yokoi}\ \emph {et~al.}(2021)\citenamefont {Yokoi},
  \citenamefont {Ma}, \citenamefont {Kasahara}, \citenamefont {Kasahara},
  \citenamefont {Shibauchi}, \citenamefont {Kurita}, \citenamefont {Tanaka},
  \citenamefont {Nasu}, \citenamefont {Motome}, \citenamefont {Hickey},
  \citenamefont {Trebst},\ and\ \citenamefont {Matsuda}}]{Yokoi2021}%
  \BibitemOpen
  \bibfield  {author} {\bibinfo {author} {\bibfnamefont {T.}~\bibnamefont
  {Yokoi}}, \bibinfo {author} {\bibfnamefont {S.}~\bibnamefont {Ma}}, \bibinfo
  {author} {\bibfnamefont {Y.}~\bibnamefont {Kasahara}}, \bibinfo {author}
  {\bibfnamefont {S.}~\bibnamefont {Kasahara}}, \bibinfo {author}
  {\bibfnamefont {T.}~\bibnamefont {Shibauchi}}, \bibinfo {author}
  {\bibfnamefont {N.}~\bibnamefont {Kurita}}, \bibinfo {author} {\bibfnamefont
  {H.}~\bibnamefont {Tanaka}}, \bibinfo {author} {\bibfnamefont
  {J.}~\bibnamefont {Nasu}}, \bibinfo {author} {\bibfnamefont {Y.}~\bibnamefont
  {Motome}}, \bibinfo {author} {\bibfnamefont {C.}~\bibnamefont {Hickey}},
  \bibinfo {author} {\bibfnamefont {S.}~\bibnamefont {Trebst}}, \ and\ \bibinfo
  {author} {\bibfnamefont {Y.}~\bibnamefont {Matsuda}},\ }\href {\doibase
  10.1126/science.aay5551} {\bibfield  {journal} {\bibinfo  {journal}
  {Science}\ }\textbf {\bibinfo {volume} {373}},\ \bibinfo {pages} {568}
  (\bibinfo {year} {2021})}\BibitemShut {NoStop}%
\bibitem [{\citenamefont {Winter}\ \emph {et~al.}(2016)\citenamefont {Winter},
  \citenamefont {Li}, \citenamefont {Jeschke},\ and\ \citenamefont
  {Valent\'{\i}}}]{Winter2016}%
  \BibitemOpen
  \bibfield  {author} {\bibinfo {author} {\bibfnamefont {S.~M.}\ \bibnamefont
  {Winter}}, \bibinfo {author} {\bibfnamefont {Y.}~\bibnamefont {Li}}, \bibinfo
  {author} {\bibfnamefont {H.~O.}\ \bibnamefont {Jeschke}}, \ and\ \bibinfo
  {author} {\bibfnamefont {R.}~\bibnamefont {Valent\'{\i}}},\ }\href {\doibase
  10.1103/PhysRevB.93.214431} {\bibfield  {journal} {\bibinfo  {journal} {Phys.
  Rev. B}\ }\textbf {\bibinfo {volume} {93}},\ \bibinfo {pages} {214431}
  (\bibinfo {year} {2016})}\BibitemShut {NoStop}%
\bibitem [{\citenamefont {Rau}\ \emph {et~al.}(2014)\citenamefont {Rau},
  \citenamefont {Lee},\ and\ \citenamefont {Kee}}]{rau2014generic}%
  \BibitemOpen
  \bibfield  {author} {\bibinfo {author} {\bibfnamefont {J.~G.}\ \bibnamefont
  {Rau}}, \bibinfo {author} {\bibfnamefont {E.~K.-H.}\ \bibnamefont {Lee}}, \
  and\ \bibinfo {author} {\bibfnamefont {H.-Y.}\ \bibnamefont {Kee}},\ }\href
  {http://journals.aps.org/prl/abstract/10.1103/PhysRevLett.112.077204}
  {\bibfield  {journal} {\bibinfo  {journal} {Phys. Rev. Lett.}\ }\textbf
  {\bibinfo {volume} {112}},\ \bibinfo {pages} {077204} (\bibinfo {year}
  {2014})}\BibitemShut {NoStop}%
\bibitem [{\citenamefont {Salavati}\ \emph {et~al.}(2019)\citenamefont
  {Salavati}, \citenamefont {Alajlan},\ and\ \citenamefont
  {Rabczuk}}]{Salavati2019}%
  \BibitemOpen
  \bibfield  {author} {\bibinfo {author} {\bibfnamefont {M.}~\bibnamefont
  {Salavati}}, \bibinfo {author} {\bibfnamefont {N.}~\bibnamefont {Alajlan}}, \
  and\ \bibinfo {author} {\bibfnamefont {T.}~\bibnamefont {Rabczuk}},\ }\href
  {\doibase https://doi.org/10.1016/j.physe.2019.05.011} {\bibfield  {journal}
  {\bibinfo  {journal} {Physica E: Low-dimensional Systems and Nanostructures}\
  }\textbf {\bibinfo {volume} {113}},\ \bibinfo {pages} {79} (\bibinfo {year}
  {2019})}\BibitemShut {NoStop}%
\bibitem [{\citenamefont {Ersan}\ \emph {et~al.}(2019)\citenamefont {Ersan},
  \citenamefont {Vatansever}, \citenamefont {Sarikurt}, \citenamefont
  {Yüksel}, \citenamefont {Kadioglu}, \citenamefont {Ozaydin}, \citenamefont
  {Üzengi Aktürk}, \citenamefont {Ümit Akıncı},\ and\ \citenamefont
  {Aktürk}}]{Ersan2019}%
  \BibitemOpen
  \bibfield  {author} {\bibinfo {author} {\bibfnamefont {F.}~\bibnamefont
  {Ersan}}, \bibinfo {author} {\bibfnamefont {E.}~\bibnamefont {Vatansever}},
  \bibinfo {author} {\bibfnamefont {S.}~\bibnamefont {Sarikurt}}, \bibinfo
  {author} {\bibfnamefont {Y.}~\bibnamefont {Yüksel}}, \bibinfo {author}
  {\bibfnamefont {Y.}~\bibnamefont {Kadioglu}}, \bibinfo {author}
  {\bibfnamefont {H.~D.}\ \bibnamefont {Ozaydin}}, \bibinfo {author}
  {\bibfnamefont {O.}~\bibnamefont {Üzengi Aktürk}}, \bibinfo {author}
  {\bibnamefont {Ümit Akıncı}}, \ and\ \bibinfo {author} {\bibfnamefont
  {E.}~\bibnamefont {Aktürk}},\ }\href {\doibase
  https://doi.org/10.1016/j.jmmm.2018.12.032} {\bibfield  {journal} {\bibinfo
  {journal} {Journal of Magnetism and Magnetic Materials}\ }\textbf {\bibinfo
  {volume} {476}},\ \bibinfo {pages} {111} (\bibinfo {year}
  {2019})}\BibitemShut {NoStop}%
\bibitem [{\citenamefont {Imai}\ \emph {et~al.}(2021)\citenamefont {Imai},
  \citenamefont {Nawa}, \citenamefont {Shimizu}, \citenamefont {Yamada},
  \citenamefont {Fujihara}, \citenamefont {Aoyama}, \citenamefont {Takahashi},
  \citenamefont {Okuyama}, \citenamefont {Ohashi}, \citenamefont {Hagihala},
  \citenamefont {Torii}, \citenamefont {Morikawa}, \citenamefont {Terauchi},
  \citenamefont {Kawamata}, \citenamefont {Kato}, \citenamefont {Gotou},
  \citenamefont {Itoh}, \citenamefont {Sato},\ and\ \citenamefont
  {Ohgushi}}]{Imai2021}%
  \BibitemOpen
  \bibfield  {author} {\bibinfo {author} {\bibfnamefont {Y.}~\bibnamefont
  {Imai}}, \bibinfo {author} {\bibfnamefont {K.}~\bibnamefont {Nawa}}, \bibinfo
  {author} {\bibfnamefont {Y.}~\bibnamefont {Shimizu}}, \bibinfo {author}
  {\bibfnamefont {W.}~\bibnamefont {Yamada}}, \bibinfo {author} {\bibfnamefont
  {H.}~\bibnamefont {Fujihara}}, \bibinfo {author} {\bibfnamefont
  {T.}~\bibnamefont {Aoyama}}, \bibinfo {author} {\bibfnamefont
  {R.}~\bibnamefont {Takahashi}}, \bibinfo {author} {\bibfnamefont
  {D.}~\bibnamefont {Okuyama}}, \bibinfo {author} {\bibfnamefont
  {T.}~\bibnamefont {Ohashi}}, \bibinfo {author} {\bibfnamefont
  {M.}~\bibnamefont {Hagihala}}, \bibinfo {author} {\bibfnamefont
  {S.}~\bibnamefont {Torii}}, \bibinfo {author} {\bibfnamefont
  {D.}~\bibnamefont {Morikawa}}, \bibinfo {author} {\bibfnamefont
  {M.}~\bibnamefont {Terauchi}}, \bibinfo {author} {\bibfnamefont
  {T.}~\bibnamefont {Kawamata}}, \bibinfo {author} {\bibfnamefont
  {M.}~\bibnamefont {Kato}}, \bibinfo {author} {\bibfnamefont {H.}~\bibnamefont
  {Gotou}}, \bibinfo {author} {\bibfnamefont {M.}~\bibnamefont {Itoh}},
  \bibinfo {author} {\bibfnamefont {T.~J.}\ \bibnamefont {Sato}}, \ and\
  \bibinfo {author} {\bibfnamefont {K.}~\bibnamefont {Ohgushi}},\ }\href@noop
  {} {\enquote {\bibinfo {title} {Magnetism of kitaev spin-liquid candidate
  material rubr$_3$},}\ } (\bibinfo {year} {2021}),\ \Eprint
  {http://arxiv.org/abs/2109.00129} {arXiv:2109.00129 [cond-mat.str-el]}
  \BibitemShut {NoStop}%
\bibitem [{\citenamefont {Ni}\ \emph {et~al.}(2021)\citenamefont {Ni},
  \citenamefont {Gui}, \citenamefont {Powderly},\ and\ \citenamefont
  {Cava}}]{Ni2021}%
  \BibitemOpen
  \bibfield  {author} {\bibinfo {author} {\bibfnamefont {D.}~\bibnamefont
  {Ni}}, \bibinfo {author} {\bibfnamefont {X.}~\bibnamefont {Gui}}, \bibinfo
  {author} {\bibfnamefont {K.~M.}\ \bibnamefont {Powderly}}, \ and\ \bibinfo
  {author} {\bibfnamefont {R.~J.}\ \bibnamefont {Cava}},\ }\href@noop {}
  {\bibfield  {journal} {\bibinfo  {journal} {arXiv preprint arXiv:2108.12915}\
  } (\bibinfo {year} {2021})}\BibitemShut {NoStop}%
\bibitem [{\citenamefont {Zhang}\ \emph {et~al.}(2021)\citenamefont {Zhang},
  \citenamefont {Lin}, \citenamefont {Moreo},\ and\ \citenamefont
  {Dagotto}}]{Zhang2021}%
  \BibitemOpen
  \bibfield  {author} {\bibinfo {author} {\bibfnamefont {Y.}~\bibnamefont
  {Zhang}}, \bibinfo {author} {\bibfnamefont {L.-F.}\ \bibnamefont {Lin}},
  \bibinfo {author} {\bibfnamefont {A.}~\bibnamefont {Moreo}}, \ and\ \bibinfo
  {author} {\bibfnamefont {E.}~\bibnamefont {Dagotto}},\ }\href@noop {}
  {\bibfield  {journal} {\bibinfo  {journal} {arXiv preprint arXiv:2111.04560}\
  } (\bibinfo {year} {2021})}\BibitemShut {NoStop}%
\bibitem [{\citenamefont {Kresse}\ and\ \citenamefont {Hafner}(1993)}]{VASP1}%
  \BibitemOpen
  \bibfield  {author} {\bibinfo {author} {\bibfnamefont {G.}~\bibnamefont
  {Kresse}}\ and\ \bibinfo {author} {\bibfnamefont {J.}~\bibnamefont
  {Hafner}},\ }\href {\doibase 10.1103/PhysRevB.47.558} {\bibfield  {journal}
  {\bibinfo  {journal} {Phys. Rev. B}\ }\textbf {\bibinfo {volume} {47}},\
  \bibinfo {pages} {558} (\bibinfo {year} {1993})}\BibitemShut {NoStop}%
\bibitem [{\citenamefont {Kresse}\ and\ \citenamefont
  {Furthm\"uller}(1996)}]{VASP2}%
  \BibitemOpen
  \bibfield  {author} {\bibinfo {author} {\bibfnamefont {G.}~\bibnamefont
  {Kresse}}\ and\ \bibinfo {author} {\bibfnamefont {J.}~\bibnamefont
  {Furthm\"uller}},\ }\href {\doibase 10.1103/PhysRevB.54.11169} {\bibfield
  {journal} {\bibinfo  {journal} {Phys. Rev. B}\ }\textbf {\bibinfo {volume}
  {54}},\ \bibinfo {pages} {11169} (\bibinfo {year} {1996})}\BibitemShut
  {NoStop}%
\bibitem [{\citenamefont {Sun}\ \emph {et~al.}(2015)\citenamefont {Sun},
  \citenamefont {Ruzsinszky},\ and\ \citenamefont {Perdew}}]{SCAN}%
  \BibitemOpen
  \bibfield  {author} {\bibinfo {author} {\bibfnamefont {J.}~\bibnamefont
  {Sun}}, \bibinfo {author} {\bibfnamefont {A.}~\bibnamefont {Ruzsinszky}}, \
  and\ \bibinfo {author} {\bibfnamefont {J.~P.}\ \bibnamefont {Perdew}},\
  }\href {\doibase 10.1103/PhysRevLett.115.036402} {\bibfield  {journal}
  {\bibinfo  {journal} {Phys. Rev. Lett.}\ }\textbf {\bibinfo {volume} {115}},\
  \bibinfo {pages} {036402} (\bibinfo {year} {2015})}\BibitemShut {NoStop}%
\bibitem [{\citenamefont {Perdew}\ \emph {et~al.}(2008)\citenamefont {Perdew},
  \citenamefont {Ruzsinszky}, \citenamefont {Csonka}, \citenamefont {Vydrov},
  \citenamefont {Scuseria}, \citenamefont {Constantin}, \citenamefont {Zhou},\
  and\ \citenamefont {Burke}}]{PBEsol}%
  \BibitemOpen
  \bibfield  {author} {\bibinfo {author} {\bibfnamefont {J.~P.}\ \bibnamefont
  {Perdew}}, \bibinfo {author} {\bibfnamefont {A.}~\bibnamefont {Ruzsinszky}},
  \bibinfo {author} {\bibfnamefont {G.~I.}\ \bibnamefont {Csonka}}, \bibinfo
  {author} {\bibfnamefont {O.~A.}\ \bibnamefont {Vydrov}}, \bibinfo {author}
  {\bibfnamefont {G.~E.}\ \bibnamefont {Scuseria}}, \bibinfo {author}
  {\bibfnamefont {L.~A.}\ \bibnamefont {Constantin}}, \bibinfo {author}
  {\bibfnamefont {X.}~\bibnamefont {Zhou}}, \ and\ \bibinfo {author}
  {\bibfnamefont {K.}~\bibnamefont {Burke}},\ }\href {\doibase
  10.1103/PhysRevLett.100.136406} {\bibfield  {journal} {\bibinfo  {journal}
  {Phys. Rev. Lett.}\ }\textbf {\bibinfo {volume} {100}},\ \bibinfo {pages}
  {136406} (\bibinfo {year} {2008})}\BibitemShut {NoStop}%
\bibitem [{\citenamefont {Ozaki}()}]{openmx}%
  \BibitemOpen
  \bibfield  {author} {\bibinfo {author} {\bibfnamefont {T.}~\bibnamefont
  {Ozaki}},\ }\href {http://link.aps.org/doi/10.1103/PhysRevB.67.155108}
  {\bibfield  {journal} {\bibinfo  {journal} {Phys. Rev. B}\ }\textbf {\bibinfo
  {volume} {67}},\ \bibinfo {pages} {155108}}\BibitemShut {NoStop}%
\bibitem [{\citenamefont {Han}\ \emph {et~al.}(2006)\citenamefont {Han},
  \citenamefont {Ozaki},\ and\ \citenamefont {Yu}}]{han2006n}%
  \BibitemOpen
  \bibfield  {author} {\bibinfo {author} {\bibfnamefont {M.~J.}\ \bibnamefont
  {Han}}, \bibinfo {author} {\bibfnamefont {T.}~\bibnamefont {Ozaki}}, \ and\
  \bibinfo {author} {\bibfnamefont {J.}~\bibnamefont {Yu}},\ }\href
  {http://link.aps.org/doi/10.1103/PhysRevB.73.045110} {\bibfield  {journal}
  {\bibinfo  {journal} {Phys. Rev. B}\ }\textbf {\bibinfo {volume} {73}},\
  \bibinfo {pages} {045110} (\bibinfo {year} {2006})}\BibitemShut {NoStop}%
\bibitem [{\citenamefont {Marzari}\ and\ \citenamefont
  {Vanderbilt}(1997)}]{MLWF1}%
  \BibitemOpen
  \bibfield  {author} {\bibinfo {author} {\bibfnamefont {N.}~\bibnamefont
  {Marzari}}\ and\ \bibinfo {author} {\bibfnamefont {D.}~\bibnamefont
  {Vanderbilt}},\ }\href {\doibase 10.1103/PhysRevB.56.12847} {\bibfield
  {journal} {\bibinfo  {journal} {Phys. Rev. B}\ }\textbf {\bibinfo {volume}
  {56}},\ \bibinfo {pages} {12847} (\bibinfo {year} {1997})}\BibitemShut
  {NoStop}%
\bibitem [{\citenamefont {Souza}\ \emph {et~al.}(2001)\citenamefont {Souza},
  \citenamefont {Marzari},\ and\ \citenamefont {Vanderbilt}}]{MLWF2}%
  \BibitemOpen
  \bibfield  {author} {\bibinfo {author} {\bibfnamefont {I.}~\bibnamefont
  {Souza}}, \bibinfo {author} {\bibfnamefont {N.}~\bibnamefont {Marzari}}, \
  and\ \bibinfo {author} {\bibfnamefont {D.}~\bibnamefont {Vanderbilt}},\
  }\href {\doibase 10.1103/PhysRevB.65.035109} {\bibfield  {journal} {\bibinfo
  {journal} {Phys. Rev. B}\ }\textbf {\bibinfo {volume} {65}},\ \bibinfo
  {pages} {035109} (\bibinfo {year} {2001})}\BibitemShut {NoStop}%
\bibitem [{\citenamefont {Weng}\ \emph {et~al.}(2009)\citenamefont {Weng},
  \citenamefont {Ozaki},\ and\ \citenamefont {Terakura}}]{Weng2009}%
  \BibitemOpen
  \bibfield  {author} {\bibinfo {author} {\bibfnamefont {H.}~\bibnamefont
  {Weng}}, \bibinfo {author} {\bibfnamefont {T.}~\bibnamefont {Ozaki}}, \ and\
  \bibinfo {author} {\bibfnamefont {K.}~\bibnamefont {Terakura}},\ }\href
  {\doibase 10.1103/PhysRevB.79.235118} {\bibfield  {journal} {\bibinfo
  {journal} {Phys. Rev. B}\ }\textbf {\bibinfo {volume} {79}},\ \bibinfo
  {pages} {235118} (\bibinfo {year} {2009})}\BibitemShut {NoStop}%
\bibitem [{\citenamefont {Dudarev}\ \emph {et~al.}(1998)\citenamefont
  {Dudarev}, \citenamefont {Botton}, \citenamefont {Savrasov}, \citenamefont
  {Humphreys},\ and\ \citenamefont {Sutton}}]{Dudarev}%
  \BibitemOpen
  \bibfield  {author} {\bibinfo {author} {\bibfnamefont {S.~L.}\ \bibnamefont
  {Dudarev}}, \bibinfo {author} {\bibfnamefont {G.~A.}\ \bibnamefont {Botton}},
  \bibinfo {author} {\bibfnamefont {S.~Y.}\ \bibnamefont {Savrasov}}, \bibinfo
  {author} {\bibfnamefont {C.~J.}\ \bibnamefont {Humphreys}}, \ and\ \bibinfo
  {author} {\bibfnamefont {A.~P.}\ \bibnamefont {Sutton}},\ }\href {\doibase
  10.1103/PhysRevB.57.1505} {\bibfield  {journal} {\bibinfo  {journal} {Phys.
  Rev. B}\ }\textbf {\bibinfo {volume} {57}},\ \bibinfo {pages} {1505}
  (\bibinfo {year} {1998})}\BibitemShut {NoStop}%
\bibitem [{\citenamefont {Ceperley}\ and\ \citenamefont {Alder}(1980)}]{CA}%
  \BibitemOpen
  \bibfield  {author} {\bibinfo {author} {\bibfnamefont {D.~M.}\ \bibnamefont
  {Ceperley}}\ and\ \bibinfo {author} {\bibfnamefont {B.~J.}\ \bibnamefont
  {Alder}},\ }\href {\doibase 10.1103/PhysRevLett.45.566} {\bibfield  {journal}
  {\bibinfo  {journal} {Phys. Rev. Lett.}\ }\textbf {\bibinfo {volume} {45}},\
  \bibinfo {pages} {566} (\bibinfo {year} {1980})}\BibitemShut {NoStop}%
\bibitem [{\citenamefont {Perdew}\ and\ \citenamefont {Zunger}(1981)}]{PZ}%
  \BibitemOpen
  \bibfield  {author} {\bibinfo {author} {\bibfnamefont {J.~P.}\ \bibnamefont
  {Perdew}}\ and\ \bibinfo {author} {\bibfnamefont {A.}~\bibnamefont
  {Zunger}},\ }\href {\doibase 10.1103/PhysRevB.23.5048} {\bibfield  {journal}
  {\bibinfo  {journal} {Phys. Rev. B}\ }\textbf {\bibinfo {volume} {23}},\
  \bibinfo {pages} {5048} (\bibinfo {year} {1981})}\BibitemShut {NoStop}%
\bibitem [{\citenamefont {Johnson}\ \emph {et~al.}(2015)\citenamefont
  {Johnson}, \citenamefont {Williams}, \citenamefont {Haghighirad},
  \citenamefont {Singleton}, \citenamefont {Zapf}, \citenamefont {Manuel},
  \citenamefont {Mazin}, \citenamefont {Li}, \citenamefont {Jeschke},
  \citenamefont {Valent\'{\i}},\ and\ \citenamefont {Coldea}}]{Johnson-arXiv}%
  \BibitemOpen
  \bibfield  {author} {\bibinfo {author} {\bibfnamefont {R.~D.}\ \bibnamefont
  {Johnson}}, \bibinfo {author} {\bibfnamefont {S.~C.}\ \bibnamefont
  {Williams}}, \bibinfo {author} {\bibfnamefont {A.~A.}\ \bibnamefont
  {Haghighirad}}, \bibinfo {author} {\bibfnamefont {J.}~\bibnamefont
  {Singleton}}, \bibinfo {author} {\bibfnamefont {V.}~\bibnamefont {Zapf}},
  \bibinfo {author} {\bibfnamefont {P.}~\bibnamefont {Manuel}}, \bibinfo
  {author} {\bibfnamefont {I.~I.}\ \bibnamefont {Mazin}}, \bibinfo {author}
  {\bibfnamefont {Y.}~\bibnamefont {Li}}, \bibinfo {author} {\bibfnamefont
  {H.~O.}\ \bibnamefont {Jeschke}}, \bibinfo {author} {\bibfnamefont
  {R.}~\bibnamefont {Valent\'{\i}}}, \ and\ \bibinfo {author} {\bibfnamefont
  {R.}~\bibnamefont {Coldea}},\ }\href {\doibase 10.1103/PhysRevB.92.235119}
  {\bibfield  {journal} {\bibinfo  {journal} {Phys. Rev. B}\ }\textbf {\bibinfo
  {volume} {92}},\ \bibinfo {pages} {235119} (\bibinfo {year}
  {2015})}\BibitemShut {NoStop}%
\bibitem [{\citenamefont {Catuneanu}\ \emph {et~al.}(2016)\citenamefont
  {Catuneanu}, \citenamefont {Kim}, \citenamefont {Can},\ and\ \citenamefont
  {Kee}}]{Catuneanu2016}%
  \BibitemOpen
  \bibfield  {author} {\bibinfo {author} {\bibfnamefont {A.}~\bibnamefont
  {Catuneanu}}, \bibinfo {author} {\bibfnamefont {H.-S.}\ \bibnamefont {Kim}},
  \bibinfo {author} {\bibfnamefont {O.}~\bibnamefont {Can}}, \ and\ \bibinfo
  {author} {\bibfnamefont {H.-Y.}\ \bibnamefont {Kee}},\ }\href {\doibase
  10.1103/PhysRevB.94.121118} {\bibfield  {journal} {\bibinfo  {journal} {Phys.
  Rev. B}\ }\textbf {\bibinfo {volume} {94}},\ \bibinfo {pages} {121118}
  (\bibinfo {year} {2016})}\BibitemShut {NoStop}%
\bibitem [{\citenamefont {Rau}\ and\ \citenamefont {Kee}(2014)}]{rau_trigonal}%
  \BibitemOpen
  \bibfield  {author} {\bibinfo {author} {\bibfnamefont {J.~G.}\ \bibnamefont
  {Rau}}\ and\ \bibinfo {author} {\bibfnamefont {H.-Y.}\ \bibnamefont {Kee}},\
  }\href {http://arxiv.org/abs/1408.4811} {\bibfield  {journal} {\bibinfo
  {journal} {arXiv preprint arXiv:1408.4811}\ } (\bibinfo {year}
  {2014})}\BibitemShut {NoStop}%
\end{thebibliography}%

\end{document}